\documentclass[conference]{IEEEtran}
\IEEEoverridecommandlockouts

{}
{}
{}

\usepackage[numbers,sort&compress,square]{natbib}
\usepackage{siunitx}
\usepackage{textcomp}
\usepackage{array}
\usepackage{nomencl}

\usepackage{amssymb}
\makenomenclature
\DeclareUnicodeCharacter{2500}{\textemdash}

\usepackage{comment}
\usepackage{etoolbox}
\usepackage{lettrine}

\usepackage{color}
\usepackage{graphicx} 
\usepackage{soul}

\usepackage{physics}

\begin{document}
\title{A Peer-to-Peer Energy Management Solution for Maximum Social Welfare}

\author{\IEEEauthorblockN{ Atefeh Alirezazadeh, Vahid Disfani}
\IEEEauthorblockA{\textit{ConnectSmart Research Laboratory, University of Tennessee at Chattanooga, Chattanooga, TN, USA} \\
 Emails: wjn166@mocs.utc.edu, vahid-disfani@utc.edu}
}

\maketitle

\begin{abstract}
\looseness=-1 In smart energy communities, prosumers who both generate and consume energy play a crucial role in shaping energy management strategies. These communities use advanced platforms that enable prosumers to actively engage in the local electricity markets by setting and adjusting their own energy prices. Through peer-to-peer (P2P) energy trading systems, members can directly exchange energy derived from sources such as solar photovoltaic panels, electric vehicle battery storage, and demand response (DR) programs. This direct exchange not only enhances the efficiency of the network but also fosters a dynamic energy market within the community. In this article, parking-sharing services for EVs and the mechanisms of P2P energy scheduling, which facilitates the transfer and communication of power among different energy communities (ECs) are addressed. It focuses on integrating solar power, responsive electrical loads, and electric vehicles (EVs) to optimize both economic returns and social benefits for all participants. The system is designed to ensure that all energy transactions are transparent and beneficial to the proactive consumers involved. Moreover, due to urban traffic conditions and the challenges of finding suitable locations for EV charging and parking, houses in these communities provide parking-sharing services for EVs. This integration of energy management and urban scheduling illustrates a holistic approach to addressing both energy and transportation challenges, ultimately leading to more sustainable urban environments.
\end{abstract}

\begin{IEEEkeywords}
Energy community, P2P energy scheduling, hosting parking, energy management, social welfare.
\end{IEEEkeywords}

\section{Introduction}\label{sec1}
\lettrine{A} new economic platform in an Energy Community (EC) revolves around the exchange of energy among consumers who generate their own power. This system, known as P2P, allows these self-generating consumers to use their solar-generated electricity for their own use, store it, or trade it with others, thereby enhancing community welfare. Typically, these prosumers have the capability to buy electricity from the main grid and sell any excess back to it. However, the compensation they receive from feed-in tariffs for selling electricity back to the grid is often significantly lower than the prices they can obtain through P2P energy exchanges \cite{1,2,3}. The P2P model offers a platform where consumers can initially engage in trading amongst themselves at local market prices before potentially trading with a retailer. These local prices typically range between retail and export prices, enabling consumers to earn revenue through the P2P energy exchange, whether they are selling or buying power. A key advantage of the P2P energy exchange is that it eliminates the need to transport power generated from green energy sources, thereby reducing transmission losses and operational costs of the power grid. Furthermore, the P2P model enhances the flexibility and resilience of the electric grid, making it better equipped to handle power outages \cite{4,5,6}. 

The P2P-based electricity market facilitates decarbonization through the use of green energy resources and storage technologies. It enhances grid efficiency, ensures reliable trading, provides voltage support, and manages congestion effectively within the system. Additionally, this model helps decrease uncertainty and minimizes losses in the local power grid. Several studies have examined the pricing mechanisms associated with P2P energy sharing. A detailed review of P2P energy sharing and trading, various trading pricing models, and the modeling tools used for P2P energy trading platforms in practical applications is presented in \cite{7}. P2P energy trading and optimal planning for multiple microgrids connected to the grid have been described in \cite{8}. This research also introduces a game theory approach to design the suggested model for multi-objective optimization to determine the optimal size of distributed energy resources and achieve optimal efficiency values.

In addition to generating income through selling photovoltaic power, EC prosumers can also earn revenue by participating in DR programs and hosting parking for EVs. The organization of EV parking is managed by the EC's aggregator. Various studies have explored the integration of EVs and DR programs. In \cite{9}, an optimization algorithm is proposed to minimize electricity costs for smart homes by scheduling smart appliances in conjunction with DR and P2P trading, employing a smart bidding strategy. Furthermore, a new method for exchanging power both with the utility and among neighbors is outlined in \cite{10}, focusing on microgrids that incorporate EVs, energy storage systems, and DR programs. In this model, microgrids operate independently in the market to maximize profits, and the market interactions eventually stabilize at a Nash equilibrium point.

This paper introduces a P2P energy scheduling solution designed to optimize power transactions among prosumers within an energy community and with the utility, aiming to maximize both welfare and profit. These prosumers engage in DR programs, operate their own EV chargers, and offer these chargers for use by other community members. The contributions of this article are outlined as follows:

1) Introducing an innovative model for P2P energy scheduling and parking-sharing services for EVs.

2) The development of a new energy management strategy that allows prosumers to engage in DR programs and photovoltaic (PV) trading, focused on maximizing revenue and profit, reducing costs, and enhancing social welfare.

3) The evaluation of the proposed solutions through Monte-Carlo simulations and the use of the K-means scenario reduction algorithm.

\begin{figure*}[t]
    \centering{\includegraphics[scale=0.21]{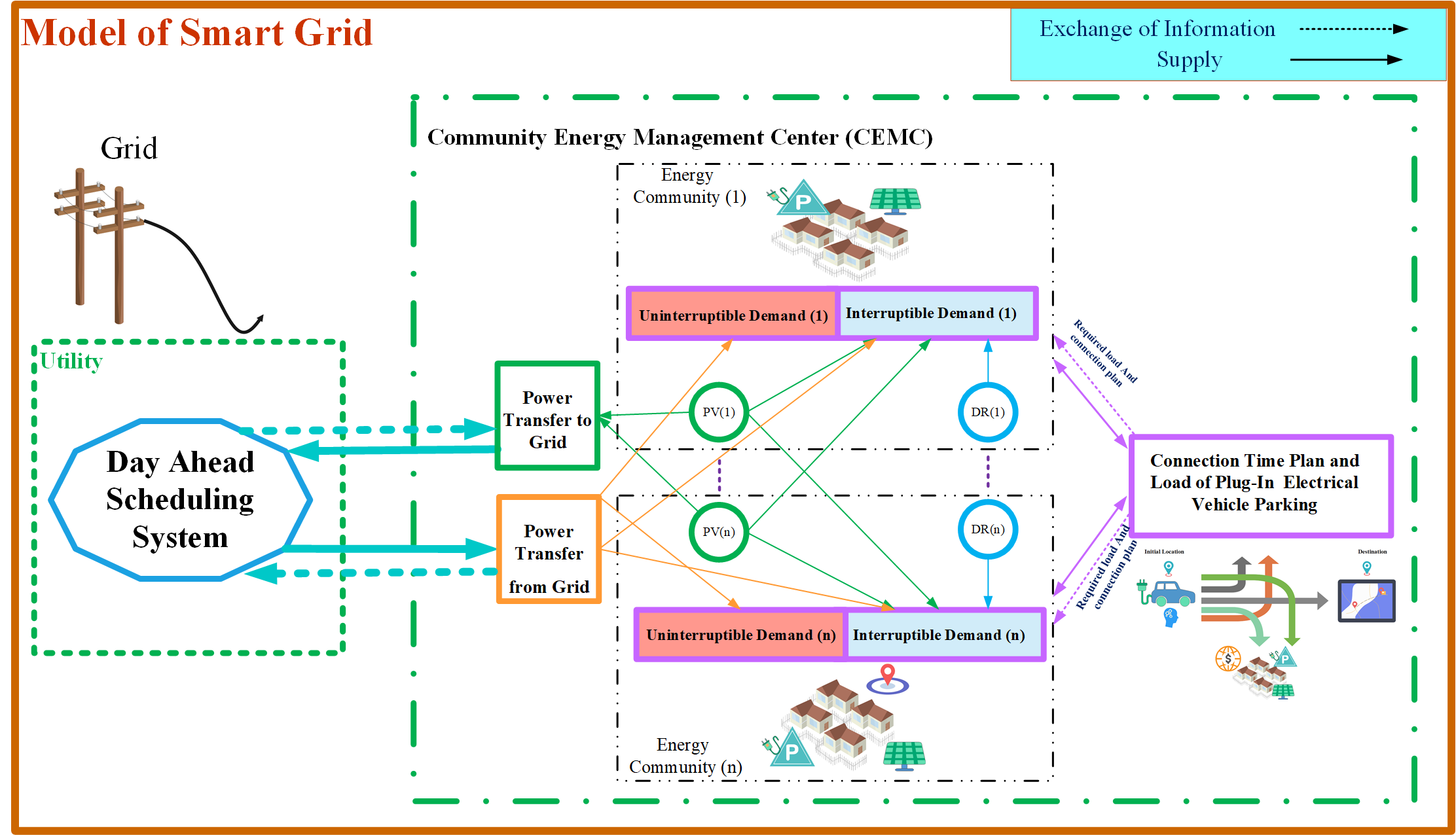}}
    \caption{The proposed model for exchange power in smart grid }\label{fig2}
\end{figure*}

\section{Approach Overview}\label{sec2}
Fig.\ref{fig2} illustrates the proposed P2P energy scheduling model within the smart Energy Community (EC). The Community Energy Management Center (CEMC) plans the system in collaboration with the utility, considering the utility's energy price, the community's internal generation, and its own power demands.

Within the CEMC, system loads are categorized into two types: uninterruptible and interruptible. Uninterruptible loads are consistently powered directly from the utility. In contrast, the interruptible loads are managed based on the EC's internal solar power generation, the internal exchanges between ECs within the P2P system or through DR programs, with any remaining required power supplied directly from the utility.

Guest EVs are responsible for paying the parking fees directly to each EC, and they charge and discharge power directly from and to the utility, settling payments accordingly. The CEMC participates in the market by sending power purchase requests or potential power selling offers to the utility based on the internal P2P market prices and the Day Ahead market price, thereby facilitating the formation of an inter-system exchange power market.

\section{Mathematical modeling}\label{sec3}
\subsection{Objective Function and Formulation}\label{sec3.1}
The optimization objective function is to maximize the total revenue of building unit ($TPC_{c,u,t}$) that is defined as:

\begin{align}[t]\label{eq1}
  & \text{ Objective Function=}\,\sum\limits_{c\in C}{\sum\limits_{u\in U}{\sum\limits_{t\in T}{\left[ \text{\textit{TPC}}_{c,u,t} \right]}}}= \nonumber \\ 
  & \sum\limits_{c\in C}{\sum\limits_{u\in U}{\sum\limits_{t\in T}{\left[
   \begin{aligned}
     &\underbrace{TPG_{c,u,t}}_{i}+\underbrace{TPDR_{c,u,t} }_{ii}+\underbrace{TPP2P_{c,u,t}}_{iii}\\
     &+\underbrace{TPSW_{c,u,t}}_{iv} +\underbrace{TPHP_{c,u,t}}_{v}+\underbrace{SCPU_{c,u,t}}_{vi} \\
     &-\underbrace{TENSIC_{c,u,t}}_{vii}+\underbrace{\left( \frac{\sum\limits_{ev\in EV}{TPEV}_{c,t,ev}}{\sum\limits_{u\in U}{\sum\limits_{ev\in EV}{{{N}_{c,t,ev}}}}}  \right)}_{viii}
    \end{aligned}
    \right]}}}
\end{align}

Part (i), the benefit from the power exchange between the EC and the utility, is expressed as follows:
\begin{equation}\label{eq2}
TPG_{c,u,t}=(PT2G_{c,u,t} - PTFG_{c,u,t}) \times c_t^G\
 \end{equation}
In this equation, $PT2G_{c,u,t}$, $PTFG_{c,u,t}$ and $c_t^G$ respectively represent the power sold by each EC to the utility, the power purchased from the utility by each EC per hour and the cost per kilowatt-hour(kWh) of energy exchange with the utility.

Part (ii), the profit or penalty resulting from the participation of ECs in DR programs, is expressed as follows:
\begin{equation}\label{eq3}
  {TPDR}_{c,u,t}{=}{DR}_{c,u,t}\times c_{t}^{DR}+({DR}_{c,u,t}-ID_{c,u,t}^{Call})\times c_{t}^{DR,Fine}
\end{equation}
In this equation, ${DR}_{c,u,t}$, $ID_{c,u,t}^{Call}$, $c_{t}^{DR}$, and $c_{t}^{DR,Fine}$ respectively, power reduced by each EC, callable power for curtailment by each EC, the income from the reduction of each kWh energy and the penalty from the non-participation of callable each kWh for curtailment per hour.

Part (iii), the benefit of neighboring ECs from inter-community energy exchange, is described as follows:
 \begin{equation}\label{eq4}
\begin{split}
{TPP2P}_{c,u,t}{=}{PT2P2P}_{c,u,t}\times c_{c,t}^{P2P}{-} {PTFP2P}_{c,u,t}\times c_{c,t}^{P2P}
 \end{split}
\end{equation}
 In this equation, ${PT2P2P}_{c,u,t}$, ${PTFP2P}_{c,u,t}$, and $c_{c,t}^{P2P}$ are respectively the power sold by each community to other communities, the power purchased by each community from other communities, and the exchange cost of each kWh of power between communities.

Part (iv), the benefit resulting from the creation of social welfare, is shown as follows:

\begin{equation}\label{eq5}
\begin{split}
   {TPSW}_{c,u,t}=(P_{c,u,t}^{Self\,\,Supply}+{PT2P2P}_{c,u,t})\times c_{c,u,t}^{SWV}
 \end{split}
\end{equation}

In this equation, $P_{c,u,t}^{Self\,\,Supply}$, ${PT2P2P}_{c,u,t}$, and $c_{c,u,t}^{SWV}$ are respectively the power provided by each EC as self-supply, the power sold to other ECs, and the value of the social welfare benefit created per kWh.

Part (v), the income from allocating the private parking of EC to EVs which are applicants for the use of EC charging station as a charging and parking place, is indicated as follows:
\begin{equation}\label{eq6}
\begin{split}
& {TPHP}_{c,u,t}= \sum\limits_{ev\in EV}{{{N}_{c,u,t,ev}}}\times c_{c,u,t}^{Parking\,\,Host}
 \end{split}
\end{equation}
In this equation, ${N}_{c,u,t,ev}$ and $c_{c,u,t}^{Parking\,\,Host}$ are the number of applicant vehicles parked in each EC and the hourly cost of parking, respectively.

Section (vi) represents the income from the standby right of ECs to participate in DR programs ($SCPU_{c,u,t}$).

Part (vii), the penalty in case of non-supply power of ECs, is shown as follows:
\begin{equation}\label{eq7}
\begin{split}
{TENSIC}_{c,u,t}={ENS}_{c,u,t}\times c_{t}^{ENS}
 \end{split}
\end{equation}
In this equation, ${ENS}_{c,u,t}$ and $c_{t}^{ENS}$ are the unsupplied power in each building unit and the hourly cost of unsupplied power, respectively.

Part (viii), the benefit from charging and discharging EVs participating in power scheduling, is shown as follows:
\begin{equation}\label{eq8}
\begin{split}
& {TPEV}_{c,t,ev}{=}{CHEV}_{c,t,ev}\times c_{t}^{CHEV} \\
& -{DCHEV}_{c,t,ev}\times c_{t}^{DCHEV}
 \end{split}
\end{equation}
In this equation, ${DCHEV}_{c,t,ev}$, ${CHEV}_{c,t,ev}$, $c_{t}^{DCHEV}$ and $C_{t}^{CHEV}$ are EV power discharging, EV power charging, hourly price of EV discharging and hourly price of EV charging, respectively.
\subsection{Constraints of the model}\label{sec3.2}
\subsubsection{Load constraints of EC }\label{sec3.2.1}

The balance constraint of power generation and consumption in the EC is stated as follows:
\begin{equation}\label{eq9}
\begin{split}
 & {ENS}_{c,u,t}=[ID_{c,u,t}^{Subsequent\ PV}-{DR}_{c,u,t}-\\
 & {PTIDFG}_{c,u,t}- {PTFP2P}_{c,u,t} ]
 \end{split}
\end{equation}
In this equation, $ID_{c,u,t}^{Subsequent\ PV}$ and ${PTIDFG}_{c,u,t}$ are the interruptible load after supply by PV and the interruptible demand purchased from the utility, respectively.

The constraints of self-supply power are shown as follows:
\begin{equation}\label{eq10}
\begin{split}
 ID_{c,u,t}^{ Subsequent \ PV }={ID}_{c,u,t}-P_{c,u,t}^{Self\,\,Supply}
 \end{split}
\end{equation}
\begin{equation}\label{eq11}
\begin{split}
 P_{c,u,t}^{Self\_Supply}\le P_{c,u,t}^{PV}-{DR}_{c,u,t}
 \end{split}
\end{equation}
which in equation (\ref{eq10}) shows the amount of interruptible power (${ID}_{c,u,t}$) after supplying power through the internal sources of the EC. Equation (\ref{eq11}) shows that the maximum self-supply power of each EC is equal to the remaining solar power generated ($P_{c,u,t}^{PV}$) after supplying the interrupted power in the DR program of that EC.

The constraints related to power exchange with the utility are shown as follows:
\begin{equation}\label{eq12}
\begin{split}
{PTUIDFG}_{c,u,t}={UID}_{c,u,t}+\frac{\sum\limits_{ev\in EV}{CHEV}_{c,t,ev}}{\sum\limits_{u\in U}{\sum\limits_{ev\in EV}{{{N}_{c,t,ev}}}}  }
 \end{split}
\end{equation}
In this equation, ${PTUIDFG}_{c,u,t}$, ${UID}_{c,u,t}$ and ${CHEV}_{c,t,ev}$ are the uninterruptible demand purchased from the utility,  the uninterruptible demand each building and the charging of the guest EV, respectively.

The maximum power that can be transmitted from the utility in the interruptible load section of each EC is expressed as follows:
\begin{equation}\label{eq13}
\begin{split}
{PTIDFG}_{c,u,t} \le (1-{v}_{c,u,t})\times {ID}_{c,u,t} ,\ \ {v}_{c,u,t} \in  \{ 0,1\}
 \end{split}
\end{equation}

The interruptible load requested from the utility is equal to the remaining interruptible load of the EC after executing the DR program, the internal supply of the EC, and the supply by other ECs that is shown as follows:
\begin{equation}\label{eq14}
\begin{split}
 & {PTIDFG}_{c,u,t} = ID_{c,u,t}^{Subsequent \ DR}-P_{c,u,t}^{Self\,\,Supply} \\
 &-{PTFP2P}_{c,u,t}
 \end{split}
\end{equation}
In this equation, $ID_{c,u,t}^{Subsequent \ DR}$ is the interruptible load after applying DR.

The total requested power of each EC from the utility is shown as follows:
\begin{equation}\label{eq15}
\begin{split}
{PTFG}_{c,u,t}={PTUIDFG}_{c,u,t}+{PTIDFG}_{c,u,t}
 \end{split}
\end{equation}

The maximum power that can be transferred to the utility is shown as follows:
\begin{equation}\label{eq16}
\begin{split}
{PT2G}_{c,u,t} \le {v}_{c,u,t}\times ( P_{c,u,t}^{PV}+\frac{\sum\limits_{ev\in EV}{DCHEV}_{c,t,ev}}{\sum\limits_{u\in U}{\sum\limits_{ev\in EV}{{N}_{c,t,ev}}}} ) 
 \end{split}
\end{equation}
In this equation $DCHEV_{c,t,ev}$ is the discharge of the guest EVs in that EC.

\subsubsection{DR program constraints of EC }\label{sec3.2.2}

The constraints related to the DR program are shown as follows:
\begin{equation}\label{eq17}
\begin{split}
{DR}_{c,u,t}\le \,ID_{c,u,t}^{Subsequent \ PV}
 \end{split}
\end{equation}
\begin{equation}\label{eq18}
\begin{split}
ID_{c,u,t}^{Subsequent \ DR}={ID}_{c,u,t}-{DR}_{c,u,t}
 \end{split}
\end{equation}

\subsubsection{P2P constraints of EC }\label{sec3.2.3}
The constraints in the exchange power of ECs are presented as follows:
\begin{equation}\label{eq19}
\begin{split}
{PT2P2P}_{c,u,t}\le P_{c,u,t}^{PV}-P_{c,u,t}^{Self \ Supply}
 \end{split}
\end{equation}
\begin{equation}\label{eq20}
\begin{split}
{PTFP2P}_{c,u,t}\le \,ID_{c,u,t}^{Subsequent \ PV}-{DR}_{c,u,t}
 \end{split}
\end{equation}
\begin{equation}\label{eq21}
\begin{split}
\sum\limits_{c\in C,u\in U}{PT2P2P}_{c,u,t}=\sum\limits_{c\in C,u\in U}{PTFP2P}_{c,u,t}
 \end{split}
\end{equation}
\begin{equation}\label{eq22}
\begin{split}
{PT2P2P}_{c,u,t}\le (1-{r}_{c,u,t})\times P_{c,u,t}^{PV} ,\ {r}_{c,u,t} \in  \{ 0,1\}
 \end{split}
\end{equation}
\begin{equation}\label{eq23}
\begin{split}
{PTFP2P}_{c,u,t}\le {r}_{c,u,t}\times {ID}_{c,u,t}
 \end{split}
\end{equation}

\subsubsection{EV constraints of EC }\label{sec3.2.4}
The existing constraints for the presence of guest EVs and their charging and discharging are presented as follows:

\begin{equation}\label{eq24}
\begin{split}
\sum\limits_{ev\in EV}UEV_{ev,c,t} \le 1 ,\ \forall c,t\in \tau
\end{split}
\end{equation}
\begin{equation}\label{eq25}
\begin{split}
\sum\limits_{c\in C}{UEV}_{ev,c,t}\le 1 ,\ \forall ev,t\in \tau 
\end{split}
\end{equation}
\begin{equation}\label{eq26}
\begin{split}
{UEV}_{ev,c,t} \le \overline{AR}_{ev,t}\times {VEV}_{ev,c},\ \forall ev,t\in \tau
\end{split}
\end{equation}
\begin{equation}\label{eq27}
\begin{split}
\sum\limits_{c\in C} VEV_{ev,c}\le 1 \ \forall ev
\end{split}
\end{equation}

Equation (\ref{eq24}) determines that each building unit in the EC can serve only one vehicle. In this equation ${UEV}_{ev,c,t} \in  \{ 0,1\} $ is a binary variable denoting the connection of each EV to each building at time t and $\tau =\left[ t_{ev}^{Con},t_{ev}^{DCon} \right]$. Equation (\ref{eq25}) shows that each vehicle can be connected to only one EC per hour. Equation (\ref{eq26}) checks the connection status of the EV ($\overline{AR}_{ev,t}$) to the desired community parking according to the previously sent request. The equation (\ref{eq27}) states that each community can serve only one vehicle during the day.
 In this equation $VEV_{ev,c}\in  \{ 0,1\} $ is a binary variable denoting the service of each community to each EV.

\section{Case Study}\label{sec4}

Based on the mathematical models described earlier, simulations have been conducted to examine the interactions between various components of an Energy Community (EC) and the utility using a sample system. This system comprises three ECs, each containing six houses. The energy consumption in each house is divided into two categories \cite{11}.  The first category is the uninterruptible load, which is directly supplied by the utility. This load typically includes everyday household usage and charging of the homeowners' EVs. The second category is the interruptible load, which can be covered by solar power generation \cite{12} within each house or through solar generation sources from P2P energy exchanges and can create income for each member of the EC by participating in DR programs. Any remaining interruptible load can be supplied by the utility if it is not met by internal sources. For each kWh that is either consumed within a house or sold to adjacent houses, there is a hidden value for the community's social welfare. The value attributed to social welfare for each kWh is considered to be 0.055 $\$$.

\begin{figure}[htp]
    \begin{minipage}[h]{0.47\linewidth}
         \centering
         \includegraphics[width=1\linewidth]{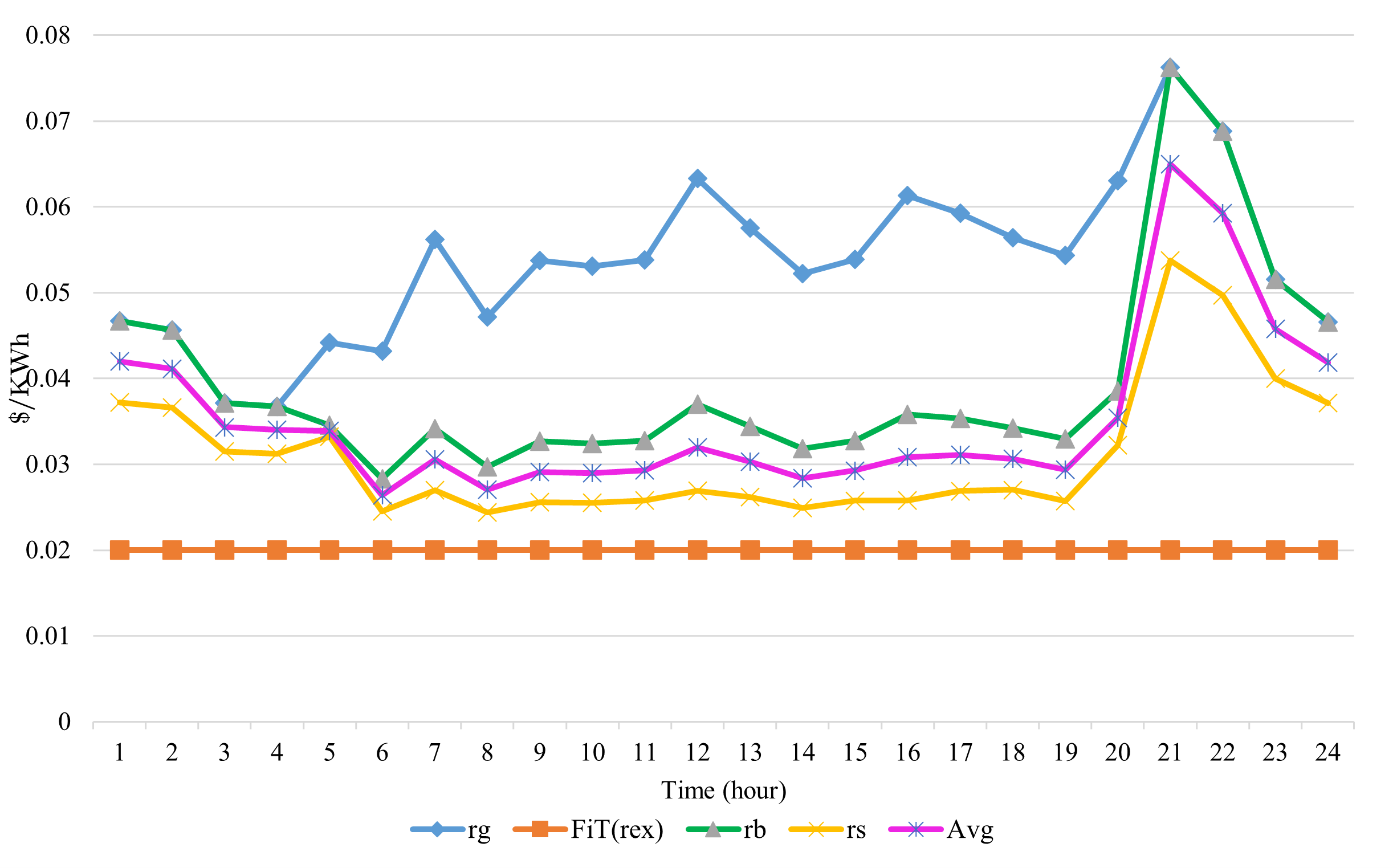}
         \caption{P2P price in electricity market}
         \label{fig9}
    \end{minipage}
    \hfill
    \vspace{0.1 cm}
    \begin{minipage}[h]{0.47\linewidth}
         \centering
         \includegraphics[width=1\linewidth]{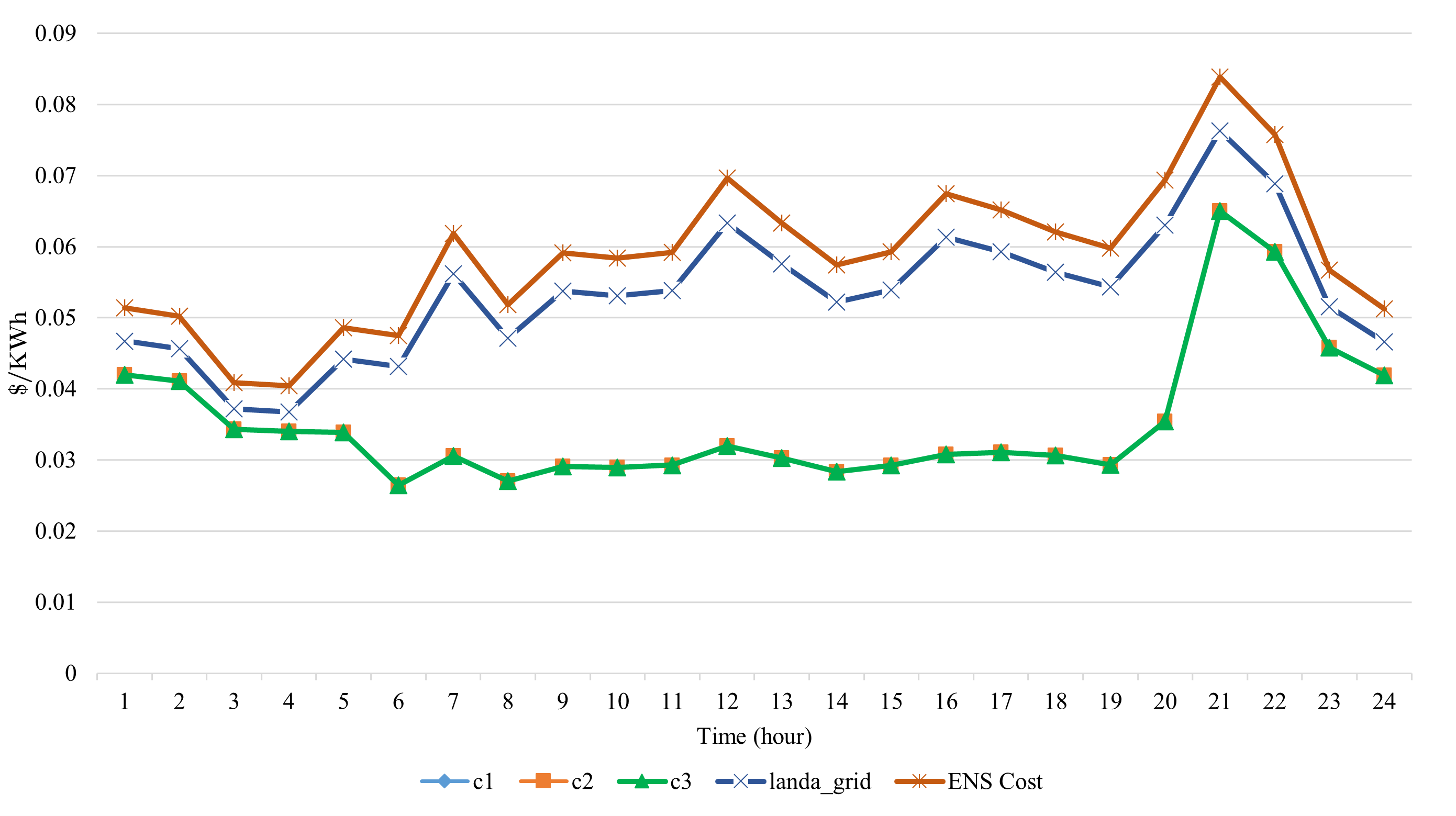}
         \caption{Energy price of utility and P2P}
         \label{fig10}     
    \end{minipage}
    \hfill
    \vspace{0.1 cm}
    \begin{minipage}[h]{0.47\linewidth}
        \centering
         \includegraphics[width=1\linewidth]{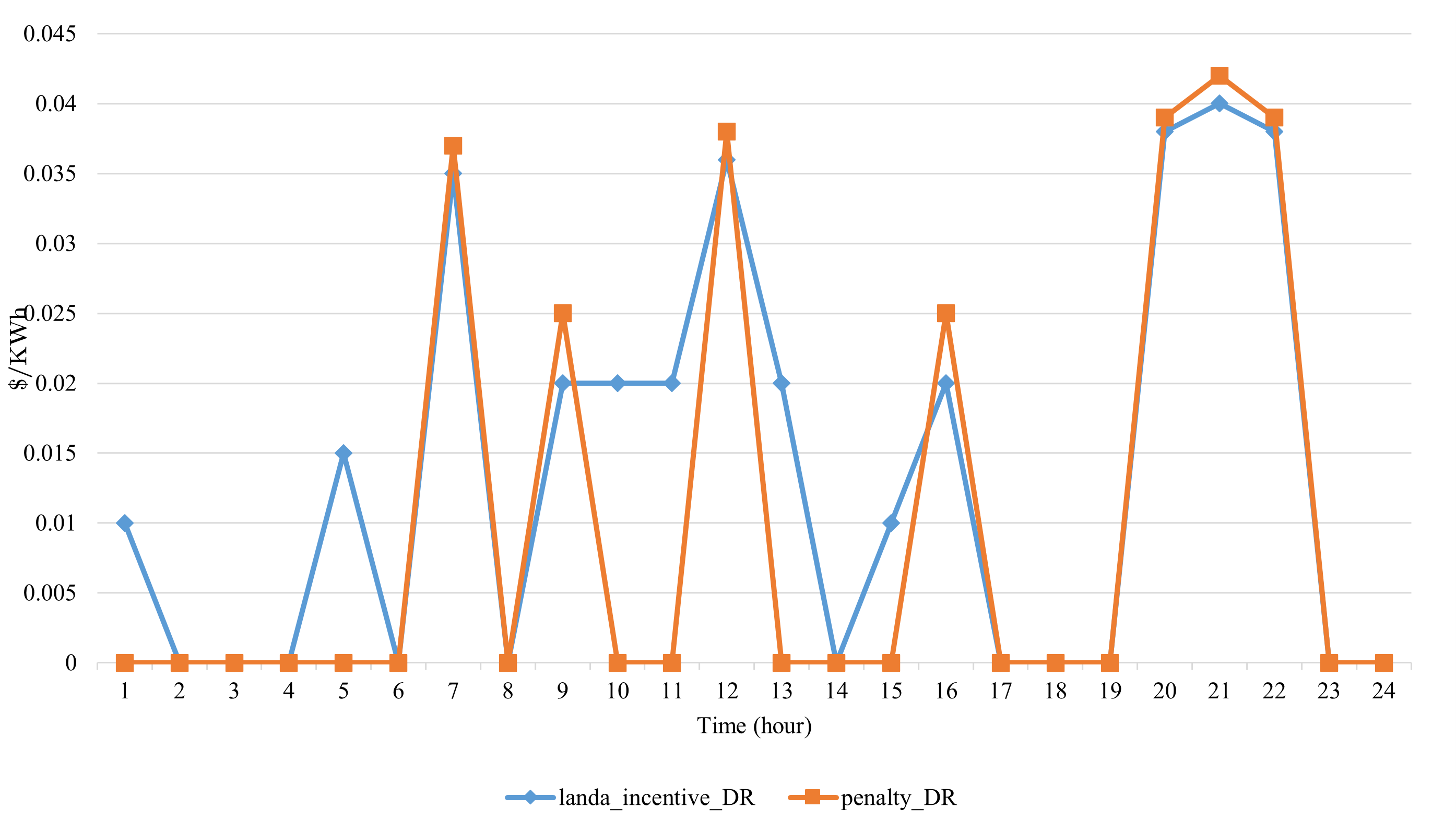}
         \caption{Incentive and penalty prices of DR programs}
         \label{fig11}
    \end{minipage}
    \hfill
    \vspace{0.1 cm}
    \begin{minipage}[h]{0.47\linewidth}
         \centering
         \includegraphics[width=1\linewidth]{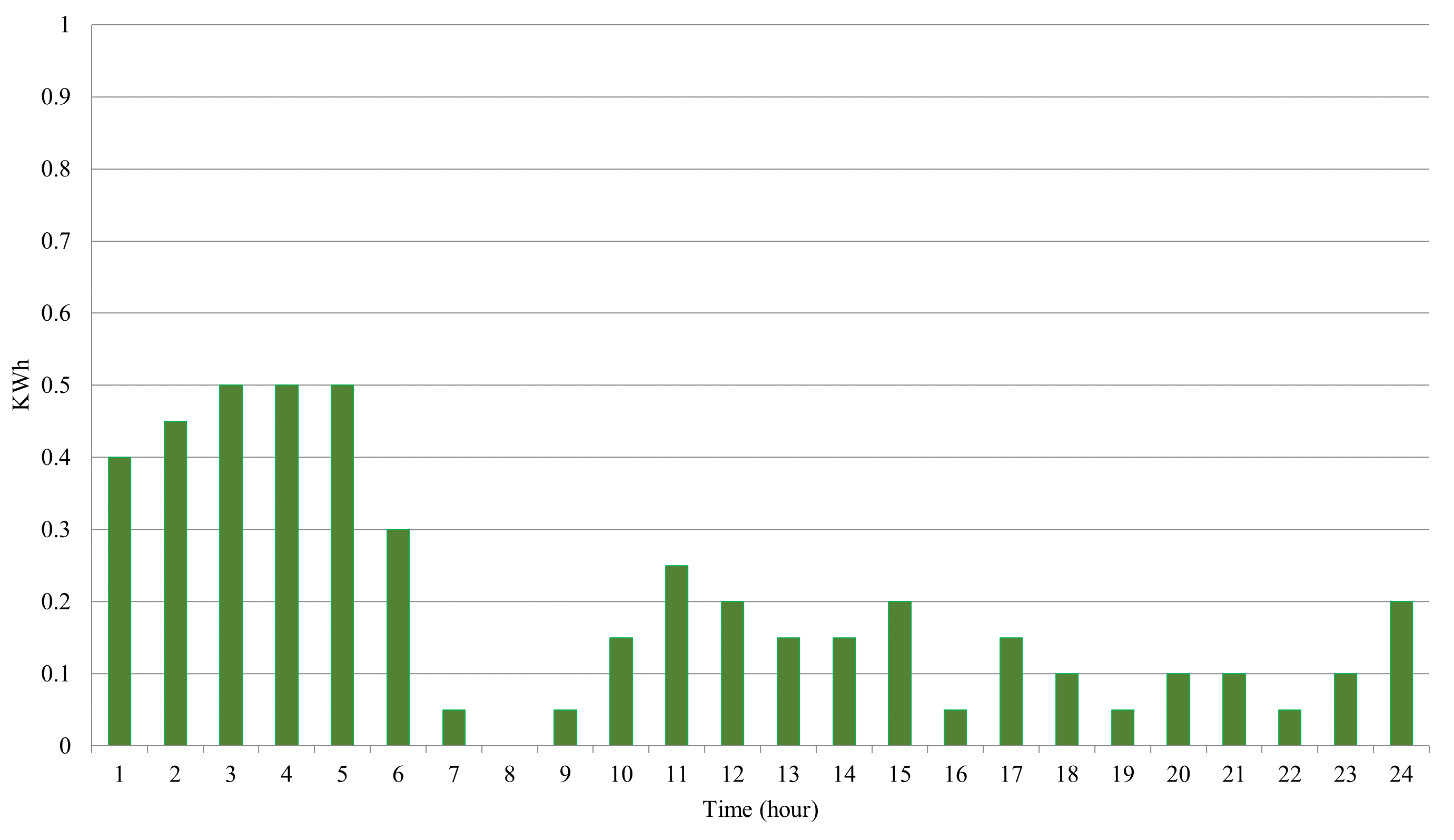}
         \caption{Allowed load to participate in DR programs}
         \label{fig12}
    \end{minipage}     
    
\end{figure}

To determine the energy exchange price among neighbors, the methodology outlined in \cite{13} is used, and the purchasing and selling prices for P2P energy are obtained as shown in Fig. \ref{fig9}. The average of these prices is considered as the P2P equilibrium price. Fig. \ref{fig10} illustrates the fundamental price of power exchange from the utility, the price of P2P power exchange, and the price of lost energy (which is set at 10$\%$ above the utility price). Additionally, the incentive and penalty rates for participants in DR programs are proposed to consumers, based on Fig. \ref{fig11}.

\begin{figure}[htp]
    \begin{minipage}[h]{0.47\linewidth}
         \centering
         \includegraphics[width=1\linewidth]{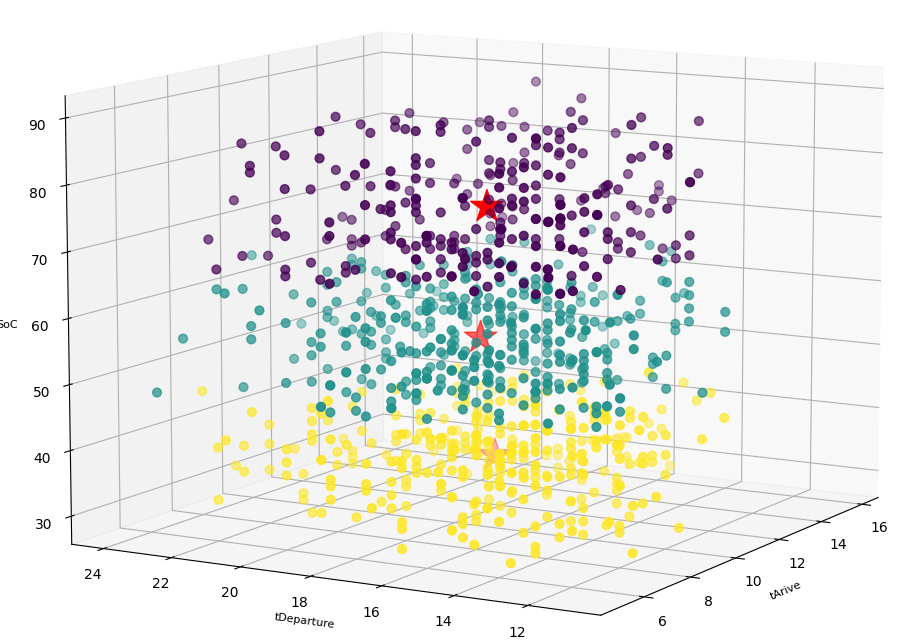}
         \caption{Scenarios reduction by K-Means algorithm}
         \label{fig14}
    \end{minipage}
    \hfill
    \vspace{0.1 cm}
    \begin{minipage}[h]{0.47\linewidth}
         \centering
         \includegraphics[width=1\linewidth]{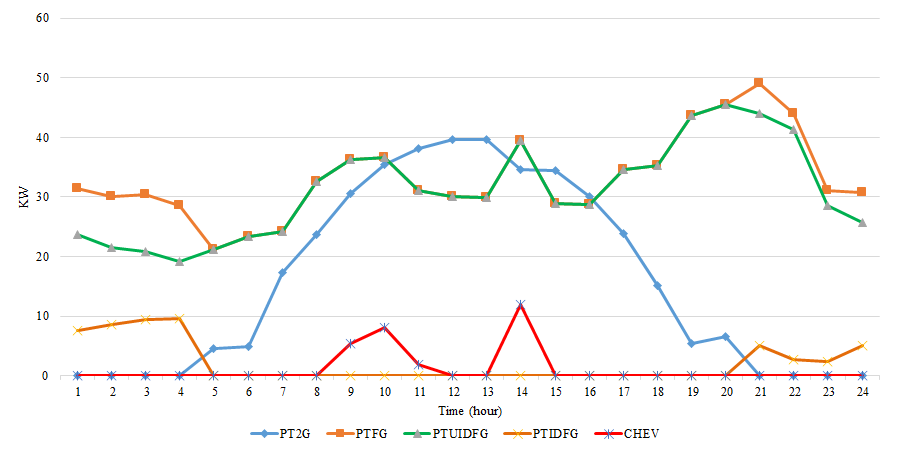}
         \caption{Exchange power transfer from/to grid}
         \label{fig16}
    \end{minipage}
    
\end{figure}

The amount of load allowed to participate in the DR program hourly is depicted in Fig. \ref{fig12}. The parameters concerning the probability function of EVs are presented in \cite{14}. Using the Mount Carlo method, 1000 different scenarios have been generated, which include the amount of charge at the time of arrival to parking, the time of arrival to the parking and departure from parking. The battery capacity of the guest EVs is 25 kWh, and the final charge required for all EVs at the time of departure is 90$\%$ of their capacity. Then, using the K-Means algorithm \cite{15}, three representatives have been selected from among different scenarios, which are illustrated in Fig. \ref{fig14}.

\section{Numerical results and discussion}\label{sec5} 

After solving the optimization model, the following outcomes have been obtained. The Fig. \ref{fig16} depicts the power sold by the entire EC to the utility. This power is the amount of solar generation power remaining after the internal consumption of each EC and the purchasing and selling of power between neighbors. According to Eq.\ref{eq21}, the power sold in the houses of sellers equals the total power purchased in the houses of buyers, with no house being both a seller and buyer simultaneously. The table (\ref{tab1}) illustrates the profit obtained in various cost and revenue sections of the entire system. As can be seen in this table, all the ECs regarding their domestic solar generation, despite selling power to the utility and generating revenue will still need to pay for the purchasing of energy to the utility. Through DR programs and renting parking to guest vehicles (1 $\$$ per hour), an income for ECs will be obtained, the values of which will be 1.11 and 25 $\$$, respectively. Guest EVs located in the host houses must pay 1.45 $\$$ to charge themselves and purchase power from the utility. For houses standby per hour to participate in DR programs, the utility will pay each house 0.1 $\$$ per hour, which will result in a total of 43.2 $\$$ of income for the EC. Due to the hidden value of social welfare created in the EC, a hidden income of 1.59 $\$$ will also be created. In total, 50.23 $\$$ will be earned during the 24-hour scheduling for the EC. The obtained results demonstrate that houses participating in the introduced scheduling, in addition to scheduling their energy supply, will earn income by participating in the P2P exchange market and renting their parking to EVs.

\begin{table}[htbp]
\centering
\caption{  Economic profits of ECs}
\label{tab1}

\begin{tabular}{ll}\\
\hline
 \textbf{Profit} & \textbf{Amount($\$$)}\\
 \hline
 \text{Total Profit from Grid } & $-22.12$ \\
 \text{Total Profit from DR Programs } & $+1.11$\\
 \text{Total Profit from Parking Hosting} & $+25$\\
 \text{Total Profit of EVs Power Exchange} & $+1.45$ \\
 \text{Total Social Welfare Profit of Customers} & $+1.59$\\
 \text{Total Profit of Customers Stand-by} & $+43.2$ \\
 \hline
 \text{Total Profit ($\$$ )} & $+50.23$ \\
\end{tabular}{}
\vspace*{-1.7\baselineskip}
\end{table}

\section{Conclusion}\label{sec6}
This paper proposed a P2P energy scheduling and electricity market for different types of households equipped with PV generation and EVs and including DR programs. The results show that by considering the P2P internal electricity market among the neighbors and energy management, the internal demands of each EC will be well supplied to increase the profitability of each participant and the entire EC. Also, the freedom of choice for each of the participants who have the ability to provide for their internal needs, provide for the needs of other neighbors, or sell power to the utility will create a kind of social welfare for them the valuation of which has been determined per hour. Moreover, since EV drivers face challenges such as long charging time and inappropriate charging locations, the use of community house parking for charging and parking vehicles every hour will solve these challenges, and bring income for the owners of the EC, which will reduce the current costs of that community and reduce urban traffic. 
\bibliography{main.bib}{}
\bibliographystyle{plain}
\end{document}